\documentclass[epj]{svjour}
\usepackage[latin1]{inputenc}
\usepackage{bm}
\usepackage{multirow}
\usepackage{amssymb}
\usepackage{amsbsy}
\usepackage{amsmath}
\usepackage{stmaryrd}
\usepackage{graphicx}
\usepackage{epsfig}

\begin{document}

\title{Projection Operator Approach to Transport in Complex Single-Particle Quantum Systems}

\author{ Robin Steinigeweg\inst{1}\thanks{\emph{\email{rsteinig@uos.de}}}%
    \and Jochen Gemmer\inst{1}\thanks{\emph{\email{jgemmer@uos.de}}}%
    \and Heinz-Peter Breuer\inst{2,3}\thanks{\emph{\email{breuer@physik.uni-freiburg.de}}}%
    \and Heinz-J\"urgen Schmidt\inst{1}}%

\institute{Fachbereich Physik, Universit\"at Osnabr\"uck, Barbarastrasse 7, D-49069 Osnabr\"uck, Germany%
\and Physikalisches Institut, Universit\"at Freiburg, Hermann-Herder-Strasse 3, D-79104 Freiburg, Germany%
\and Hanse-Wissenschaftskolleg, Institute for Advanced Study, D-27753 Delmenhorst, Germany}%

\date{\today}

\abstract{We discuss the time-convolutionless (TCL) projection
operator approach to transport in closed quantum systems. The
projection onto local densities of quantities such as energy,
magnetization, particle number, etc.~yields the reduced dynamics
of the respective quantities in terms of a systematic perturbation
expansion. In particular, the lowest order contribution of this
expansion is used as a strategy for the analysis of transport in
``modular'' quantum systems corresponding to quasi one-dimensional
structures which consist of identical or similar many-level
subunits. Such modular quantum systems are demonstrated to
represent many physical situations and several examples of complex
single-particle models are analyzed in detail. For these quantum
systems lowest order TCL is shown to represent an efficient tool
which also allows to investigate the dependence of transport on
the considered length scale. To estimate the range of validity of
the obtained equations of motion we extend the standard projection
to include additional degrees of freedom which model non-Markovian
effects of higher orders.}



\maketitle

%
%

\section{Introduction}
\label{introduction} In recent years the application of projection
operator techniques
\cite{nakajima1958,zwanzig1960,chaturvedi1979,breuer2007,breuer2006,breuer2007b}
to transport investigations in (closed) quantum systems has been
suggested \cite{steinigeweg2007-1,weimer2008,wu2008}. Within these
approaches a suitable projection onto local densities of pertinent
transport quantities is used in order to obtain the reduced
dynamics of these quantities in terms of a systematic perturbation
expansion, typically w.r.t.~some (small) interaction strength. For
certain examples the lowest order contribution of this expansion
has been shown to yield reliable predictions on transport
\cite{weimer2008} as well as on its length scale dependence
\cite{steinigeweg2007-1}.

However, the lowest order truncation may become questionable for
strongly interacting quantum systems, of course. But already in
the weak coupling case the validity of this truncation generally
is restricted to short time scales \cite{bartsch2008}. This fact
seems to be problematic, especially since the relevant time scale
can be very long for small interaction strengths or large length
scales \cite{steinigeweg2007-1}, e.~g., in the thermodynamic
limit. For this reason the additional consideration of higher
order contributions appears to be indispensable \cite{breuer2006},
contrary to the statements in, e.~g., Ref.~\cite{wu2008}. Because
such a consideration has not been provided in the literature so
far, the main intention of the present paper is the incorporation
of higher order terms as well, particularly the next-to-lowest
order contribution.

To start with we will introduce the notion of a modular quantum
system in Sec.~\ref{modular}. Modular quantum systems are
demonstrated to represent many physical situations and several
examples will be given. In the context of these quantum systems
the time-convolutionless (TCL) projection operator technique
\cite{chaturvedi1979,breuer2006,breuer2007} is subsequently
discussed. In Sec.~\ref{projection} the projection onto local
densities and lowest order TCL is firstly shown as an appropriate
method which also allows to investigate the dependence of
transport on the considered length scale. In particular explicit
conditions for the applicability of the introduced method are
given. The next Sec.~\ref{extension} is concerned with the higher
order contributions of the TCL expansion and a suitable estimation
is derived for the special case of interactions with van Hove
structure \cite{vanhove1954,vanhove1957,bartsch2008}. In
Sec.~\ref{range} this estimation is used in order to determine the
range of validity of lowest order TCL and an interpretation in the
context of length scales is provided. Section \ref{application}
finally applies the concepts of the previous Sections to complex
single-particle models, e.~g., to models with disorder. Our
results are confirmed by numerical solution of the full
time-dependent Schr\"odinger equation.

%
%

\section{Modular Quantum Systems and Diffusive Dynamics}
\label{modular} In the present paper we consider so-called
``modular'' quantum systems which have a quasi one-dimensional
structure and consist of $N$ identical or at least similar
many-level subunits. These subunits are described by a local
Hamiltonian $\hat{h}_\mu$ and the next-neighbor interaction
between two adjacent subunits is denoted by $\lambda \,
\hat{v}_{\mu,\mu+1}$, where $\lambda$ adjusts the overall coupling
strength. The total Hamiltonian $\hat{H} = \hat{H}_0 + \lambda \,
\hat{V}$ is given by
\begin{equation}
\hat{H}_0 = \sum_{\mu = 0}^{N-1} \hat{h}_\mu \, , \; \hat{V} =
\sum_{\mu = 0}^{N-1} \hat{v}_{\mu, \mu+1} \, , \label{hamiltonian}
\end{equation}
where we employ periodic boundary conditions, i.~e., we identify
$\mu = N$ with $\mu = 0$. Such a description obviously applies to
one-dimensional structures such as chains of atoms, molecules,
quantum dots, etc. But also $s = 1/2$ spin chains fit into this
scheme of description, because a segment of the chain, that is, a
number of spins and their mutual interactions can be chosen in
order to form a suitable many-level subunit. Similarly, spin
lattices or higher-dimensional models of the Hubbard type can be
treated by the use of (\ref{hamiltonian}), if a whole chain (2D)
or layer (3D) is considered as a single subunit. Thus, for a large
class of quantum systems an adequate way of description is offered
by the Hamiltonian of Eq.~(\ref{hamiltonian}), see
Sec.~\ref{application} for details.

In this paper we primarily deal with ``single-particle'' quantum
systems, that is, those quantum systems which allow to restrict
the investigation to a linearly instead of an exponentially
increasing Hilbert space. This is mainly done in order to open the
possibility for a comparison of the theoretical predictions with
the numerical solution of the full time-dependent Schr\"odinger
equation. Hence, we may suppose
\begin{equation}
\hat{h}_\mu = \sum_{i = 0}^{n-1} h_\mu^i \, | \mu, i \rangle
\langle \mu , i | \, , \label{h}
\end{equation}
\begin{equation}
\hat{v}_{\mu, \mu+1} = \sum_{i, j = 0}^{n-1} v_{\mu, \mu +1}^{i,j}
\, | \mu, i \rangle \langle \mu +1 , j | + \text{H.c.} \, ,
\label{v}
\end{equation}
where $n$ denotes the number of levels within a subunit. Without
loss of generality, we have additionally assumed an off-diagonal
block structure of the interaction, that is, $| \mu, i \rangle$
and $| \mu, j \rangle$ are not coupled.

Of particular interest is the local density $p_\mu(t)$ of, e.~g.,
energy, magnetization, excitations, particles, probability, etc.
This quantity is expressed as the expectation value of a
corresponding operator $\hat{p}_\mu$,
\begin{equation}
p_\mu(t) = \text{Tr} \{ \rho(t) \, \hat{p}_\mu \} \, , \;
\hat{p}_\mu = \sum_{i=0}^{n-1} p_\mu^i \, | \mu, i \rangle \langle
\mu , i | \, , \label{p}
\end{equation}
where $\rho(t)$ is the full system's density matrix. Since the
operators $\hat{p}_\mu$ are assumed to be diagonal in the energy
representation of the uncoupled system, we restrict ourselves to
those quantities $p_\mu(t)$ which are conserved for the special
case of $\lambda = 0$. However, this restriction still allows to
investigate transport for a large class of systems, as will be
demonstrated in Sec.~\ref{application}. Our aim is to analyze the
dynamical behavior of the $p_\mu(t)$ and to develop explicit
criteria which enable a clear distinction between diffusive and
other available types of transport, e.~g., ballistic or insulating
behavior.

The dynamical behavior may be called diffusive if the $p_\mu(t)$
fulfill a discrete diffusion equation
\begin{equation}
\dot{p}_\mu(t) = D \, [ \, p_{\mu-1}(t)- 2 \, p_\mu(t) +
p_{\mu+1}(t) \, ] \label{diffusion1}
\end{equation}
with some $\mu$- and $t$-independent diffusion constant $D$. It is
straightforward to show [multiplying (\ref{diffusion1}) by $\mu$,
respectively $\mu^2$, summarizing over $\mu$ and manipulating
indices on the r.h.s.] that the spatial variance
\begin{equation}
\sigma^2(t) = \sum_{\mu = 0}^{N-1} \mu^2 \, p_\mu(t) - \left [ \,
\sum_{\mu = 0}^{N-1} \mu \, p_\mu(t) \, \right ]^2
\label{variance}
\end{equation}
increases linearly with $t$, namely $\sigma^2(t) = 2 \, D \, t$.
By contrast, ballistic behavior is characterized by $\sigma^2(t)
\propto t^2$, whereas insulating behavior corresponds to
$\sigma^2(t) = \text{const.}$

According to Fourier's work, diffusions equations are routinely
decoupled with respect to, e.~g., cosine-shape spatial density
profiles
\begin{equation}
p_q(t) = C_q \sum_{\mu = 0}^{N-1} \cos(q \, \mu) \, p_\mu(t) \, ,
\; q = \frac{2 \pi \, k}{N} \, ,
\end{equation}
with $k = 0, 1, \ldots, N/2$ and a suitable normalization constant
$C_q$. Consequently, Eq.~(\ref{diffusion1}) yields
\begin{equation}
\dot{p}_q(t) = -2 \, (1 - \cos q) \, D \, p_q(t) \, .
\label{diffusion2}
\end{equation}
Therefore, if the quantum model indeed shows diffusive transport,
all modes $p_q(t)$ have to relax exponentially. If, however, the
$p_q(t)$ are found to relay exponentially only for some regime of
$q$, the model is said to behave diffusively on the corresponding
length scale $l = 2 \pi / q$. One might think of a length scale
which is both large compared to some mean free path [below that
ballistic behavior occurs, $\sigma^2(t) \propto t^2$] and small
compared to, say, some localization length [beyond that insulating
behavior appears, $\sigma^2(t) = \text{const.}$], see
Sec.~\ref{application}.

%
%

\section{Projection onto Local Densities and Second Order TCL}
\label{projection} A strategy for the analysis of the dynamical
behavior of the $p_q(t)$ is provided by the time-convolutionless
(TCL) projection operator technique
\cite{chaturvedi1979,breuer2006,breuer2007}. This technique, and
the well-known Nakajima-Zwanzig (NZ) method as well, are applied
in order to describe the reduced dynamics of a quantum system with
a Hamiltonian of the form $\hat{H} = \hat{H}_0 + \lambda \,
\hat{V}$, see Refs.~\cite{nakajima1958,zwanzig1960}. Even though
projection operator methods are well-established approaches in the
context of open systems, the following application to transport in
closed systems is a novel concept. Let us remark that the
application of these techniques only requires that the pertinent
observables commute with $\hat{H}_0$.

Generally, the full dynamics of a quantum system is given by the
Liouville-von Neumann equation
\begin{equation}
\frac{\partial}{\partial t} \, \rho(t) = - \imath \, [ \, \lambda
\, \hat{V}(t), \rho(t) \, ] = {\cal L}(t) \, \rho(t) \, ,
\end{equation}
where time arguments refer to the interaction picture. In order to
describe the reduced dynamics of the system, one has to construct
a suitable projection operator $\cal P$ which projects onto the
relevant part of the density matrix $\rho(t)$. In particular,
$\cal P$ has to satisfy the property ${\cal P}^2 = {\cal P}$.
Because in our case the relevant variables are the local densities
$p_\mu(t)$, we choose
\begin{equation}
{\cal P} \, \rho(t) = \sum_{\mu = 0}^{N-1} \text{Tr} \{ \rho(t) \,
\hat{p}_\mu \} \, \hat{p}_\mu = \sum_{\mu = 0}^{N-1} p_\mu(t) \,
\hat{p}_\mu \, . \label{projector}
\end{equation}
This choice indeed fulfills the property ${\cal P}^2 = \cal P$, if
we additionally normalize
\begin{equation}
\text{Tr} \{ \, (\hat{p}_\mu)^2 \, \} = \sum_{i = 0}^{n-1}
(p_\mu^i)^2 = 1 \, ,
\end{equation}
which can be done without loss of generality. Note that this
normalization typically implies $p_\mu^i \sim 1 / \sqrt{n}$, at
least for the models in Sec.~\ref{application}.

Once some projection operator has been defined, the TCL formalism
routinely yields [not only for our choice of $\cal P$ in
Eq.~(\ref{projector})] a closed and time-local equation for the
dynamics of ${\cal P} \, \rho(t)$,
\begin{equation}
\frac{\partial}{\partial t} \, {\cal P} \, \rho(t) = {\cal K}(t)
\, {\cal P} \, \rho(t) + {\cal I}(t) \, (1- {\cal P}) \, \rho(0),
\label{tcl}
\end{equation}
and avoids the often troublesome time-convolution which appears,
e.~g., in the context of the NZ method. Eq.~(\ref{tcl}) and, in
particular, its time-locality are a standard result and a direct
consequence of the TCL formalism
\cite{chaturvedi1979,breuer2006,breuer2007}. The time-locality of
the TCL equation may be understood as the result of a forward and
backward propagation of the corresponding NZ equation in time. In
Sec.~\ref{application} we will briefly comment on the NZ approach
and, especially, on its implications for our class of models.

For initial conditions $\rho(0)$ with ${\cal P} \, \rho(0) =
\rho(0)$ the inhomogeneity ${\cal I}(t)$ on the r.h.s.~of
(\ref{tcl}) vanishes. [But for the models in
Sec.~\ref{application} there are numerical indications that ${\cal
I}(t)$ is even negligible for other $\rho(0)$, see also
Refs.~\cite{michel2005,breuer2006,bartsch2008}.] The generator
${\cal K}(t)$ is given as a systematic perturbation expansion in
powers of the coupling strength $\lambda$,
\begin{equation}
 {\cal K}(t) = \sum_{m = 1}^{\infty} \lambda^m \, {\cal K}_m(t) \, .
\label{generator}
\end{equation}
The odd contributions of this expansion vanish for many models and
for our models as well, that is, ${\cal K}_{2m-1}=0$.
Consequently, the lowest non-vanishing contribution is the second
order ${\cal K}_2(t)$ which reads
\begin{equation}
{\cal K}_2(t) = \int_0^t \text{d}t_1 \, {\cal P} \, {\cal L}(t) \,
{\cal L}(t_1) \, {\cal P} \, . \label{k2}
\end{equation}
The next non-vanishing contribution is the fourth order ${\cal
K}_4(t)$ which is given by
\begin{eqnarray}
{\cal K}_4(t) \! &=& \! \int_0^t \text{d}t_1 \int_0^{t_1}
\text{d}t_2 \int_0^{t_2} \text{d}t_3 \nonumber \\[0.1cm]
&& \! \Big[ {\cal P} \, {\cal L}(t) \, {\cal L}(t_1) \, (1 - {\cal
P})
\, {\cal L}(t_2) \, {\cal L}(t_3) \, {\cal P} \nonumber \\[0.1cm]
&& \! -{\cal P} \, {\cal L}(t) \, {\cal L}(t_2) \, {\cal P} \,
{\cal L}(t_1) \, {\cal L}(t_3) \, {\cal P} \nonumber \\[0.1cm]
&& \! -{\cal P} \, {\cal L}(t) \, {\cal L}(t_3) \, {\cal P} \,
{\cal L}(t_1) \, {\cal L}(t_2) \, {\cal P} \Big] \, . \label{k4}
\end{eqnarray}

The truncation of the generator (\ref{generator}) to lowest order,
i.~e., its approximation by the second order (\ref{k2}), is
commonly done in the case of small $\lambda$ but is in general
restricted to short time scales. As already mentioned in the
introduction, this truncation seems to be problematic, since the
relevant time scale can be very long, especially if $\lambda$ is
small. However, the rest of this Section is only devoted to the
second order truncation. In the following Sec.~\ref{extension} we
will additionally discuss the fourth order (\ref{k4}) and
subsequently show that its incorporation is indispensable in order
to estimate the accuracy of the second order prediction. Moreover,
we will demonstrate that some non-diffusive transport phenomena
such as localization can not be predicted correctly by a mere
second order consideration, see Sec.~\ref{application}.

Plugging the projector (\ref{projector}) into (\ref{tcl}) and
(\ref{k2}) leads to
\begin{equation}
\dot{p}_\mu(t) = \lambda^2 \!\! \sum_{\nu = \mu-1}^{\mu+1} \!\!
R_{\mu, \nu}(t) \, p_\nu(t) \, . \label{rates}
\end{equation}
Note that the sum does not run over all $\nu$, because only
adjacent subunits are coupled. The time-dependent rates $R_{\mu,
\nu}(t)$ are defined by
\begin{equation}
R_{\mu,\nu}(t) = \int_0^t \text{d}\tau \, C_{\mu, \nu}(\tau) \, ,
\label{definitionR}
\end{equation}
where we have introduced the correlation functions
\begin{equation}
C_{\mu, \nu}(\tau) = \text{Tr} \Big \{ [ \, \hat{p}_\mu,
\hat{V}(t) \, ] [ \, \hat{p}_\nu, \hat{V}(t_1) \, ] \Big \} \, ,
\; \tau = t - t_1 \label{definitionC}
\end{equation}
with $C_{\mu, \nu}(\tau) = C_{\nu, \mu}(\tau)$. [The trace is
independent from the order of commutators and time arguments.]

So far, Eq.~(\ref{rates}) is exact. For the further simplification
of (\ref{rates}) two assumptions have to be made now.
\\
({\bf i}.)~We assume that $R_{\mu, \mu-1}(t)$, $R_{\mu, \mu}(t)$
and $R_{\mu, \mu+1}(t)$ depend only negligibly on the concrete
choice of $\mu$. Hence, it follows that $R_{\mu, \mu-1}(t) \approx
R_{\mu+1, \mu}(t) = R_{\mu, \mu+1}(t)$. This is fulfilled exactly,
if the system is translational invariant, e.~g., if the
coefficients in (\ref{h}), (\ref{v}) and (\ref{p}) are the same
for each subunit.
\\
({\bf ii}.)~Moreover, we assume $R_{\mu, \mu-1}(t) \approx
-R_{\mu, \mu}(t) / 2$. This is perfectly satisfied if the sum of
all local densities is conserved since the overall conservation
implies
\begin{equation}
\sum_{\nu = 0}^{N-1} [ \, \hat{p}_\nu, \hat{V}(t_1) \, ] = 0 \, .
\end{equation}
Consequently, the sum of $C_{\mu, \nu}(\tau)$ over all $\nu$,
eventually $2 \, C_{\mu, \mu-1}(\tau) + C_{\mu, \mu}(\tau)$, also
vanishes.
\\
Due to ({\bf i}.) and ({\bf ii}.) the Fourier transform of
(\ref{rates}) reads
\begin{equation}
\dot{p}_q(t) = -W \, R(t) \, p_q(t) \, , \qquad W = 2 \, (1 - \cos
q) \, \lambda^2 \label{tcl2}
\end{equation}
with a single rate $R(t) = -R_{\mu, \mu}(t) / 2$. Note that this
rate is still time-dependent. Remarkably, the dependence on $q$
and $\lambda$ simply appears as an overall scaling factor $W$.

The models in Sec.~\ref{application} typically feature a
correlation function $C(\tau)$ which decays completely within some
time scale $\tau_C$. After this correlation time $C(\tau)$
approximately remains zero and $R(t)$ takes on a constant value
$R$, the area under the initial peak of $C(\tau)$. Since the
correlation time apparently is independent from $q$ and $\lambda$,
it is always possible to realize a relaxation time $\tau_R \propto
1/W$ which is much larger than $\tau_C$, e.~g., in an infinite
system there definitely is a small enough $q$. For $\tau_C \ll
\tau_R$ the second order truncation (\ref{tcl2}) immediately
yields
\begin{equation}
\dot{p}_q(t) = -W \, R \; p_q(t), \label{tcl2diffusion}
\end{equation}
and the comparison with (\ref{diffusion2}) clearly indicates
diffusive behavior with a diffusion constant $D = \lambda^2 \, R$.

However, the present method is not restricted to the investigation
of diffusive transport phenomena and (\ref{tcl2}) is applicable as
well in order to completely classify the dynamics of $p_q(t)$
which decay on relatively short time scales below $\tau_C$ or on
very long time scales where, e.~g., correlations may reappear, see
Ref.~\cite{steinigeweg2007-1} and Sec.~\ref{application}.

%
%

\section{Extension of the Projection and Fourth Order TCL}
\label{extension} This Section and the next Sec.~\ref{range} as
well are concerned with the question, why and to what extend the
projection onto local densities and second order TCL yield
reliable predictions for the dynamical behavior of the $p_q(t)$.
The answer to this question surely requires the consideration of
the higher order contributions of the TCL expansion, e.~g., the
investigation of the fourth order ${\cal K}_4(t)$, cf.~(\ref{k4}).
But already the direct evaluation of ${\cal K}_4(t)$ turns out to
be extremely difficult in general, both analytically and
numerically. The problem is mainly caused by the first integrand
in the expression (\ref{k4}) which significantly contributes to
${\cal K}_4(t)$.

We therefore present an alternative approach which is essentially
based on the following idea: The influence of the higher order
terms may decrease substantially, if the projection additionally
incorporates variables which are not of particular interest by
themselves but potentially affect the dynamical behavior of the
local densities. This idea is obviously only useful if the
complexity of the higher order terms decreases to a larger extend
than the complexity of the second order contribution increases.

We concretely choose $N \, (n-1)$ additional variables
$a_\mu^i(t)$ which are also given as the expectation values of
corresponding operators $\hat{a}_\mu^i$. These operators are
assumed to be diagonal in the energy representation of the
uncoupled system and are given by
\begin{equation}
\hat{a}_\mu^i = \sum_{j = 0}^{n-1} a_\mu^{i,j} \, | \mu, j \rangle
\langle \mu , j | \label{a_i}
\end{equation}
with $\mu = 0,1,\ldots,N-1$ and $i = 0,1,\ldots,n-2$.
Additionally, the operators are supposed to fulfill
\begin{equation}
\text{Tr} \{ \hat{a}_\mu^i \, \hat{a}_\mu^j \} = \delta_{i,j} \, ,
\; \text{Tr} \{ \hat{a}_\mu^i \, \hat{p}_\mu \} = 0
\end{equation}
such that the set of all $\hat{a}_\mu^i$, $\hat{p}_\mu$ spans the
whole space of diagonal matrices. The extended projector
\begin{equation}
\tilde{{\cal P}} \, \rho(t) = {\cal P} \, \rho(t) + \sum_{\mu =
0}^{N-1} \sum_{i = 0}^{n-2} \text{Tr} \{ \rho(t) \, \hat{a}_\mu^i
\} \, \hat{a}_\mu^i \label{projector2}
\end{equation}
consequently projects onto the diagonal elements of the density
matrix $\rho(t)$. Thus, the complement $1 - \tilde{\cal P}$ in the
first integrand of (\ref{k4}) is a projection onto non-diagonal
elements. But non-diagonal contributions are negligible if we
restrict ourselves to interactions with the so-called van Hove
structure, that is, $\hat{V}^2$ is essentially a diagonal matrix,
cf.~Refs.~\cite{vanhove1954,vanhove1957,bartsch2008}. (If the van
Hove property is not fullfilled, the second order prediction is
not reliable at all \cite{bartsch2008}.) The latter fact indicates
that the extended projector transforms the largest part of
original fourth order effects into the second order. For the
general case of interactions without van Hove structure, however,
the choice of diagonal operators in Eq.~(\ref{a_i}) may not
represent the most relevant variables for fourth order
corrections, of course.

Plugging the extended projector (\ref{projector2}) into
(\ref{tcl}), (\ref{k2}) [and assuming ({\bf i}.)~translational
invariance, ({\bf ii}.)~overall conservation of local densities]
one finds
\begin{eqnarray}
\!\!\! \dot{p}_\mu(t) \! &=& \! \lambda^2 \, R(t) \, [ \,
p_{\mu-1}(t) - 2 \, p_\mu(t) + p_{\mu+1}(t) \, ] \nonumber \\
\!\!\! &+& \! \lambda^2 \!\! \sum_{\nu = \mu-1}^{\mu+1} \sum_{i =
0}^{n-2} \alpha_{\mu, \nu}^i(t) \, a_{\nu}^i(t) \, ,
\label{rates1a} \\[0.2cm]
\!\!\! \dot{a}_\mu^i(t) \! &=& \! \lambda^2 \!\! \sum_{\nu =
\mu-1}^{\mu+1} \!\! \alpha_{\nu, \mu}^i(t) \, p_{\nu}(t) + \sum_{j
= 0}^{n-2} \alpha_{\mu, \nu}^{i, j}(t) \, a_\nu^j(t)
\label{rates2a}
\end{eqnarray}
with time-dependent rates $\alpha_{\mu, \nu}^{i}(t)$ and
$\alpha_{\mu, \nu}^{i, j}(t)$ which are defined analogously to
(\ref{definitionR}) as integrals over correlation functions
$C_{\mu, \nu}^i(\tau)$ and $C_{\mu, \nu}^{i, j}(\tau)$, of course.
While $C_{\mu, \nu}^i(\tau)$ is given by
\begin{equation}
C_{\mu, \nu}^i(\tau) = \text{Tr} \Big \{ [ \, \hat{p}_\mu,
\hat{V}(t) \, ] [ \, \hat{a}_\nu^i, \hat{V}(t_1) \, ] \Big \} \, ,
\label{C_i}
\end{equation}
$C_{\mu, \nu}^{i, j}(\tau)$ is obtained by using $\hat{a}_\mu^i$
and $\hat{a}_\nu^j$ as first arguments in the above commutators.
For simplicity, however, we set the corresponding rates
$\alpha_{\mu, \nu}^{i, j}(t) = 0$ for the following reason:
Because we still consider initial conditions with $a_\mu^i(0) =
0$, a significant increase of $a_\mu^i(t)$ certainly arises only
from the first part of (\ref{rates2a}), at least for sufficiently
small times.

The translational invariance implies that $\alpha_{\mu,
\mu-1}(t)$, $\alpha_{\mu, \mu}(t)$ and $\alpha_{\mu, \mu+1}(t)$ do
not depend on the concrete choice of $\mu$. (For clarity we
suppress the fixed index $i$ here.) But $\alpha_{\mu, \mu-1}(t)
\approx \alpha_{\mu, \mu+1}(t)$ only holds true if we make the
additional assumption ({\bf iii.})~$\alpha_{\mu, \nu}(t) \approx
\alpha_{\nu, \mu}(t)$, which is exactly fulfilled for mirror
symmetry, that is, the coefficients in (\ref{v}) do not depend on
the index order. The overall conservation of the local densities
finally leads to $\alpha_{\mu, \mu-1}(t) = -\alpha_{\mu, \mu}(t) /
2$.
\\
Applying the Fourier transform to (\ref{rates1a}) yields
\begin{eqnarray}
\dot{p}_q(t) = -W \Big [ \,R(t) \, p_q(t) + \sum_{i = 0}^{n-2}
\alpha^i(t) \, a_q^i(t) \, \Big ] \label{rates1b}
\end{eqnarray}
and the Fourier transform which results only from the first part
of (\ref{rates2a}) is given by
\begin{equation}
\dot{a}_q^i(t) = -W \, \alpha^i(t) \, p_q(t) \label{rates2b}
\end{equation}
with rates $\alpha^i(t) = -\alpha_{\mu, \mu}^i(t) / 2$. Finally,
integrating (\ref{rates2b}) and inserting into (\ref{rates1b})
leads to
\begin{eqnarray}
\dot{p}_q(t) \! &=& \! [\, -W \, R(t) + W^2
\, S(t) \, ] \; p_q(t) \, , \nonumber \\[0.2cm]
S(t) \! &=& \! \int_0^t \! \text{d}t_1 \, \frac{p_q(t_1)}{p_q(t)}
\, \sum_{i = 0}^{n-2} \alpha^i(t) \, \alpha^i(t_1) \, . \label{S}
\end{eqnarray}

Remarkably, the extended projection has lead to an additional
contribution with an overall scaling factor $W^2 = 4 \, (1-\cos
q)^2 \lambda^4$. This simple scaling suggests the equivalence
between small coupling strengths and large length scales.
Especially, the factor $\lambda^4$ indicates that the additional
contribution can be interpreted partially as a fourth order effect
of the original projection. Note that the neglected right part of
(\ref{rates2a}) would lead to further contributions which scale
with higher powers of $\lambda$.

In the rest of this Section we intend to further simplify the
above equation and also link the results to those which were found
in Ref.~\cite{bartsch2008}. Plugging (\ref{a_i}) into (\ref{C_i})
leads to
\begin{equation}
C^i(\tau) = \sum_{j = 0}^{n-1} a_\mu^{i,j} \, g_\mu^j(\tau) \, ,
\end{equation}
where $g_\mu^i(\tau)$ are the diagonal elements of the matrix
\begin{equation}
\hat{g}_\mu(t,t_1) = \frac{1}{2} \, [ \, \hat{V}(t), [ \,
\hat{V}(t_1), \hat{p}_\mu \, ]] \, ,
\end{equation}
that is, $g_\mu^i(\tau) = \langle i, \mu \, | \,
\hat{g}_\mu(t,t_1) \, | \, \mu ,i \rangle$. (In the following the
fixed index $\mu$ is suppressed.) We directly obtain
\begin{equation}
\sum_{k = 0}^{n-2} C^k(\tau) \, C^k(\tau_1) = \sum_{i,j = 0}^{n-1}
g^i(\tau) \, g^j(\tau_1) \sum_{k = 0}^{n-2} a^{k,i} \, a^{k,j} \,
.
\end{equation}
Since the set of all $\hat{a}^i$, $\hat{p}$ forms a complete
orthonormal basis, the $k$-sum on the r.h.s.~of the above equation
is identical to $\delta_{i,j} - p^i \, p^j$. As a consequence the
remaining sums over $i$ and $j$ can be performed independently
from each other. Finally, by integrating over the independent
variables $\tau$ and $\tau_1$, a straightforward calculation leads
to
\begin{equation}
\sum_{k = 0}^{n-2} \alpha^k(t) \, \alpha^k(t_1) = \sum_{i =
0}^{n-1} G^i(t) \, G^i(t_1) - R(t) \, R(t_1) \, ,
\end{equation}
where $G^i(t)$ is the integral corresponding to $g^i(\tau)$. Since
$p_q(t_1) \geq p_q(t)$, we eventually end up with the ``best
case'', if we simply set $p_q(t_1) = p_q(t)$ in (\ref{S}), that
is,
\begin{equation}
S(t) \geq \int_{0}^t \! \text{d}t_1 \sum_{i = 0}^{n-1} G^i(t) \,
G^i(t_1) - R(t) \, R(t_1) \, .
\end{equation}
The ``worst case'' results by setting $p_q(t_1) = p_q(0)$, e.~g.,
$p_q(t_1) / p_q(t) \leq e \approx 2.7$ for $t \leq \tau_R$.

It is worth to mention that the above equation does not depend on
the additional variables. Moreover, since we typically consider
relaxation times which are much larger than the time scale at
which correlations decay, $G^i(t_1)$ and $R(t_1)$ are
approximately constant rates, at least as long as correlations do
not reappear. We hence obtain
\begin{equation}
S(t) \geq t \left[ \, \sum_{i = 0}^{n-1} G^i(t)^2 - R(t)^2 \right
] \, . \label{approximationS}
\end{equation}
This result coincides with the fourth order estimation $S(t)$
which were derived for investigations in the context of relaxation
in closed quantum systems, see Ref.~\cite{bartsch2008}.

%
%

\section{Range of Validity of the Second Order}
\label{range} In this Section we are going to quantify the
validity range of the second order prediction which is obtained
from the original projection onto the local densities only. To
this end we will define a measure $\chi$ which is suitable for any
situation. But in the context of completely decaying and not
reappearing correlations this measure directly determines the
range of length scales on which diffusive behavior is to be
expected, that is,
\begin{equation}
\frac{q_\text{min}}{q_\text{max}} =
\frac{l_\text{min}}{l_\text{max}} \approx 2 \, \sqrt{\chi} \, ,
\label{rangel}
\end{equation}
where $q_\text{min}$ ($l_\text{max}$), $q_\text{max}$
($l_\text{min}$) correspond to the longest, respectively shortest
exponentially relaxing $p_q(t)$. Below $l_\text{min}$ ballistic
transport occurs, whereas beyond $l_\text{max}$ any non-diffusive
type of transport, e.~g., insulating behavior may emerge.

To start with, we consider the two contributions which occur in
(\ref{S}). Their ratio
\begin{equation}
f(t) = \frac{W^2 S(t)}{W \, R(t)} = W \, \frac{S(t)}{R(t)}
\label{definitionf}
\end{equation}
typically is a monotonically increasing function,
cf.~(\ref{approximationS}). As a consequence there always exists a
time $t_B$ with $f(t_B) = 1$, that is, a time where both
contributions are equally large. But this fact does not restrict
the validity of the second order prediction, if $t_B \gg \tau_R$
and therefore $f(\tau_R) \ll 1$. The validity obviously breaks
down only in the case of, say, $f(\tau_R) \approx 1$ or even
larger.

Since $f(t)$ and $\tau_R$ depend on $W$, we use the definition of
the relaxation time
\begin{equation}
\exp \! \Big [ \! -W \! \int_0^{\tau_R} \!\!\! \text{d}t_1 \,
R(t_1) \, \Big ] = \exp \! \Big[ \! -1 \Big]
\label{definitiontauR}
\end{equation}
in order to replace $W$ in (\ref{definitionf}). Due to this
replacement $f(\tau_R)$ becomes a function of the free variable
$\tau_R$,
\begin{equation}
f(\tau_R) = S(\tau_R) \, \Big [ \, R(\tau_R) \int_0^{\tau_R}
\!\!\! \text{d}t_1 \, R(t_1) \, \Big ]^{-1} \, .
\end{equation}
($\tau_R$ still depends on $W$, of course.) Because also
$f(\tau_R$) usually turns out to increase monotonically, we define
$\text{max}(\tau_R)$ as the maximum $\tau_R$ for which $f(\tau_R$)
is still smaller than $1$. Note that this maximum relaxation time
already specifies the validity range of the second order
prediction.

However, we usually deal with decaying correlations and it is
hence useful to set $\text{max}(\tau_R)$ in relation to $\tau_C$.
We therefore define the measure $\chi$ as the dimensionless
quantity $\chi = \tau_C / \text{max}(\tau_R)$. For example, $\chi
= 1$ directly implies the breakdown of the second order prediction
on relatively short time scales on the order of $\tau_C$, whereas
$\chi = 0$ strongly indicates its unrestricted validity.

For practical purposes an interpretation of $\chi$ in the context
of length scales certainly is advantageous. Such an interpretation
essentially requires the inversion of (\ref{definitiontauR}). In
general this can only be done by numerics. But if correlations
decay completely and do not reappear, we have $\tau_R = 1/(W \,
R)$ for $\tau_R \gg \tau_C$. For sufficiently small $q$ we may
approximate $W \approx q^2 \, \lambda^2$. Therefore, for fixed
$\lambda$, we may write
\begin{eqnarray}
&& \frac{1}{W_\text{max} \, R} = \frac{1}{q_\text{max}^2 \, \lambda^2
\, R} = 2 \, \tau_C \, , \nonumber \\
&& \frac{1}{W_\text{min} \, R} = \frac{1}{q_\text{min}^2 \, \lambda^2
\, R} = \frac{\text{max}(\tau_R)}{2} \, ,
\end{eqnarray}
where the factors $2$ and $1/2$ are chosen to slightly fulfill
$\tau_C \ll \tau_R \ll \text{max}(\tau_R)$, that is,
$q_\text{min}$, $q_\text{max}$ correspond to the longest,
respectively shortest exponentially relaxing $p_q(t)$. We finally
end up with $q_\text{min} / q_\text{max} \approx 2 \,
\sqrt{\chi}$. (Analogously, for fixed $q$, one obtains
$\lambda_\text{min} / \lambda_\text{max} \approx 2 \,
\sqrt{\chi}$.)

It remains to clarify what estimation for $S(t)$ should be chosen
for the calculation of $\chi$. This choice basically depends on
the intention: One may show that $\chi$ is small even in the
``worst case'' or that $\chi$ is large in spite of the ``best
case'' assumption, see Sec.~\ref{application}.

%
%

\section{Application to Models}
\label{application}

%
%

\subsection{Modular Quantum Systems with Random Interactions}
\label{random}

\begin{figure}[htb]
\centering
\includegraphics[width=0.85\linewidth]{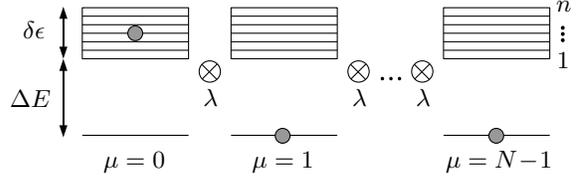}
\caption{A chain of $N$ identical, weakly coupled subunits which
feature a non-degenerate ground state, an energy gap $\Delta E$
and an energy band $\delta \epsilon$ with $n$ equidistant states.
The dots indicate excitation probabilities and are supposed to
visualize a state from the investigated ``single-excitation
space''.} \label{figuremodel}
\end{figure}

In the present Section we firstly consider a model which is
``designed'' for the application of our method, since it perfectly
fulfills almost all properties which have been assumed in the
previous Secs.~\ref{projection}-\ref{range}. The projection onto
local densities and second order TCL remarkably allow for a
complete characterization of all available types of transport and
their dependence on the considered length scale, too.

According to Fig.~\ref{figuremodel}, the model is a chain of $N$
identical subunits which are assumed to feature a non-degenerate
ground state, a large energy gap $\Delta E$ and a comparatively
narrow energy band $\delta \epsilon$ with $n$ equidistant states.
In the following we focus on the ``single-excitation subspace'',
that is, only one subunit is excited to its band, while all other
subunits are in their ground states. Consequently, the local
Hamiltonian is given by (\ref{h}) with $\mu$-independent
coefficients $h^i = i \, \delta \epsilon / n + \Delta E$.

The next-neighbor interaction (\ref{v}) is also supposed to be
identical for all adjacent subunits. In particular the
$\mu$-independent coefficients $v^{i,j}$ form a normalized matrix
whose elements are chosen at random from a Gaussian distribution
with zero mean. But we do not intend to apply random matrix theory
or discuss quantum chaos. For our purposes the crucial point is
the fact that matrices of this kind satisfy the van Hove property,
see Sec.~\ref{extension}.

However, following the ideas of quantum chaos, we expect that each
sufficiently complex one-particle system will take on a form which
is similar to our model, once it is partitioned into local
subunits and those subunits are diagonalized. The local
interactions will effectively be random (Gaussian
orthogonal/unitary ensemble) and the local spectra will be more or
less equidistant (Wigner level statistics). The influence of the
spectral details is discussed below.

Our system may be viewed as a simplified model for, e.~g., a chain
of coupled atoms, molecules, quantum dots, etc. In this case the
hopping of the excitation from one subunit to another corresponds
to transport of energy, especially if $\Delta E \gg \delta
\epsilon$. This system may also be viewed as a tight-binding model
for particles on a lattice. In this case the hopping corresponds
to transport of particles. There are $n$ ``orbitals'' per lattice
site but apparently no particle-particle interaction in the sense
of the Hubbard model. Although the $v^{i,j}$ are random, these are
systems without disorder in the sense of, say, Anderson
\cite{anderson1958}, since the $v^{i,j}$ are independent from
$\mu$. For some literature on this class of systems, see
Refs.~\cite{michel2005,gemmer2006,steinigeweg2006-2,steinigeweg2007-1,kadiroglu2007}.
Remarkably, a very similar model has been used in
Ref.~\cite{weaver2006} in order to investigate the flow of wave
energy in reverberation room acoustics.

In the context of this model we are mainly interested in the
probability $p_\mu(t)$ for finding an excitation of the $\mu$th
subunit to its band, while all other subunits are in their ground
state. This quantity is described by (\ref{p}) with
 $p^i = 1/\sqrt{n}$. Because the model is
translational invariant and the overall conservation of
probability is naturally provided, the projection onto these local
quantities and TCL2 immediately leads to (\ref{tcl2}), that is,
the relaxation of $p_q(t)$ with the standard rate $-W \, R(t)$.
The underlying correlation function $C(\tau)$ reads
\begin{equation}
C(\tau) = \frac{2}{n} \sum_{i,j = 0}^{n-1} | v^{i,j} |^2 \, \cos
\! \Big [ \, \frac{\delta \epsilon \, (i - j) \, \tau}{n} \, \Big
] \, .
\end{equation}

\begin{figure}[htb]
\centering
\includegraphics[width=0.7\linewidth]{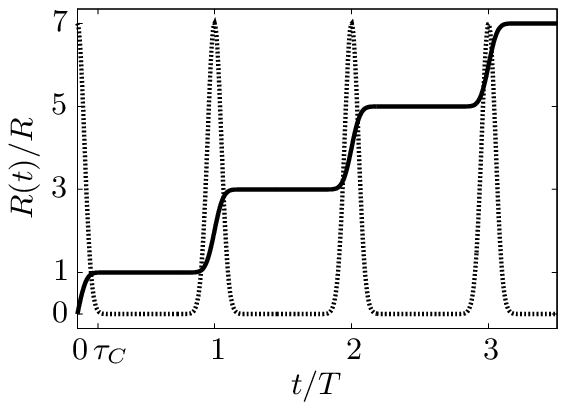}
\caption{Sketch of the correlation function $C(\tau)$ [dashed
line] and its integral $R(t)$ [continuous line]: $C(\tau)$
features complete revivals at multiples of $T$ such that $R(t)$
has a step-like form. This sketch indeed reflects the numerical
results for $C(\tau)$ and $R(t)$ but highlights the relevant time
scales for clearness. \label{figureC}}
\end{figure}

Of course, $C(\tau)$ depends on the concrete realization of the
random coefficients $v^{i,j}$. But, due to the law of large
numbers, the crucial features are nevertheless the same for the
overwhelming majority of all realizations, at least as long as
$\sqrt{n} \gg 1$. And in fact, $C(\tau)$ typically assumes the
form in Fig.~\ref{figureC}. It decays like a standard correlation
function on a time scale in the order of $\tau_C = 1 / \delta
\epsilon$. The area under this initial peak is approximately given
by $R = 2 \pi \, n / \delta \epsilon$. However, because the local
bands possess an equidistant level spacing, $C(\tau)$ is a
strictly periodic function with the period $T = 2 \pi \, n /
\delta \varepsilon$, unlike standard correlation functions. As a
consequence its time integral $R(t)$ nearly represents a step
function, see Fig.~\ref{figureC}. A non-equidistant level spacing
will certainly ``smooth'' these steps and change their width,
height and distance as well. But we expect that the tendency of an
increasing rate $R(t)$ will nevertheless remain.

\begin{figure}[htb]
\centering
\includegraphics[width=0.7\linewidth]{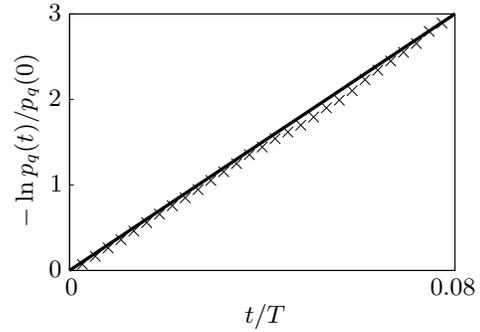}
\caption{Time evolution of a mode $p_{\pi}(t)$ which decays on an
intermediate time scale $\tau_C < t < T$. Numerics (crosses) shows
an exponential decay which indicates diffusive behavior and is in
accord with the theoretical prediction (continuous line).
Parameters: $N = 120$, $n = 500$, $\delta \epsilon = 0.5$,
$\lambda =0.0005$. \label{figuredecay1}}
\end{figure}

Along the lines of Sec.~\ref{projection}, for $\tau_C < t < T$
diffusive behavior with a diffusion constant $D = \lambda^2 \, R$
is to be expected, cf.~(\ref{tcl2diffusion}). And indeed, for
density profiles $p_q(t)$ which decay on such an intermediate time
scale we find an excellent agreement between (\ref{tcl2diffusion})
and the numerical solution of the full time-dependent
Schr\"odinger equation (which is obtained by incorporating Bloch's
theorem and exactly diagonalizing the Hamiltonian within decoupled
subspaces). In Fig.~\ref{figuredecay1} a ``typical'' example is
shown for a single realization of the random numbers $v^{i,j}$.

\begin{figure}[htb]
\centering
\includegraphics[width=0.8\linewidth]{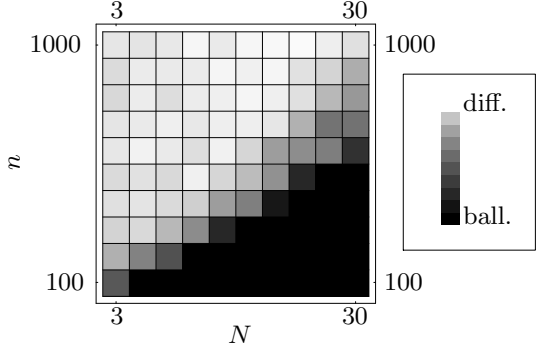}
\caption{Deviations of the time evolution of Fourier modes
$p_q(t)$ with $q = 2 \pi/N$, the longest wavelength, from a purely
exponential decay for different model parameters $N$ and $n$.
These deviations are based on a measure used in
\cite{steinigeweg2006-2} and are in accord with the claim that
diffusive transport behavior is restricted to the regime defined
by condition (\ref{condition1}). Other model parameters: $\delta
\epsilon = 0.5$, $\lambda = 0.0005$.} \label{figurephase}
\end{figure}

However, until now the above picture is not complete for two
reasons. The first reason is that $p_q(t)$ may decay on a time
scale that is long compared to $T$. According to (\ref{tcl2}),
this will happen, if $2 \, (1 - \cos q) \, \lambda^2 R \, T \gg 1$
is violated. If we approximate $2 \, (1 - \cos q) \approx q^2 = 4
\pi^2 / \, l^2$ for rather small $q$ (large $l$), we obtain the
condition
\begin{equation}
\left( \frac{4 \pi^2 \, n \, \lambda}{l \, \delta \epsilon}
\right)^2 \gg 1 \, . \label{condition1}
\end{equation}
If this condition is satisfied for the largest possible $l$,
i.~e., for $l = N$, the system exhibits diffusive behavior for all
modes. If, however, the system is large enough to allow for some
$l$ that violates condition (\ref{condition1}), diffusive behavior
breaks down in the long-wavelength limit. This result is again
backed up by numerics, see Fig.~\ref{figurephase}.

\begin{figure}[htb]
\centering
\includegraphics[width=0.8\linewidth]{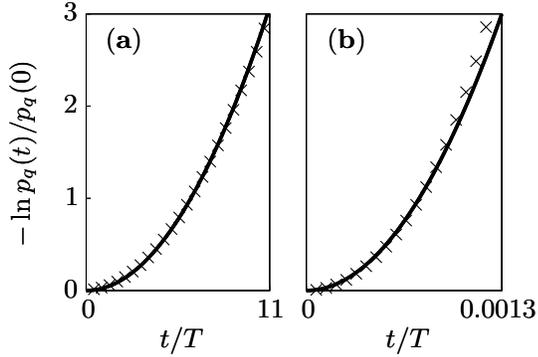}
\caption{({\bf a}) Evolution of a mode $p_{\pi/60}(t)$ which decays
on a time scale $t \gg T$, parameters: $N=120$, $n=500$, $\delta
\epsilon=0.5$, $\lambda=0.0005$. ({\bf b}) Evolution of a mode
$p_{\pi}(t)$ which decays on a time scale $t \ll \tau_C$,
parameters: $N =120$, $n=500$, $\delta \epsilon=0.5$,
$\lambda=0.004$. In both cases numerics (crosses) shows a Gaussian
decay which indicates ballistic behavior and is in accord with the
theoretical predictions (continuous lines).} \label{figuredecay2}
\end{figure}

Towards what transport type does the system deviate from
diffusive, if condition (\ref{condition1}) is violated? As already
mentioned above, we have to consider time scales $t \gg T$ in this
regime. We may thus approximate $R(t) \approx 2 \, R \, t / T$,
see Fig.~\ref{figureC}. Plugging $D(t) = 2 \, \lambda^2 R \, t /
T$ into (\ref{variance}) leads to a spatial variance $\sigma^2(t)
= 2 \, \lambda^2 R \, t^2 / T$, clearly indicating a transition
towards ballistic transport. The validity of our approach is again
backed up by numerics: In the ballistic regime a Gaussian decay of
$p_q(t)$ is to be expected, see Fig.~\ref{figuredecay2}a.

In a second case transport may be non-diffusive, if the $p_q(t)$
decay on a time scale that is short compared to $\tau_C$. This
will happen, if $2 \, (1 - \cos q) \, \lambda^2 R \, \tau_C \ll 1$
is violated. If we use the approximation $2 \, (1- \cos q) \approx
4$ for the largest possible $q$ (smallest possible $l$), the above
inequality may be written as
\begin{equation}
\frac{8 \pi \, n \, \lambda^2}{\delta \epsilon^2} \ll 1 \, .
\label{condition2}
\end{equation}
If this inequality is violated, diffusive behavior breaks down in
the limit of short-wavelength modes. Moreover, if the second order
still yields reasonable results for not too large $\lambda$, we
expect a linearly increasing rate $R(t)$ and thus a Gaussian
decay, that is, according to the above reasoning, ballistic
transport. For increasing wavelength, however, the corresponding
inequality will eventually be satisfied, hence allowing for
diffusive behavior. Also these conclusions are in accord with
numerics, see Fig.~\ref{figuredecay2}b.

%
%

\subsubsection*{Relationship to Standard Solid State Theory}

\begin{figure}[htb]
\centering
\includegraphics[width=0.7\linewidth]{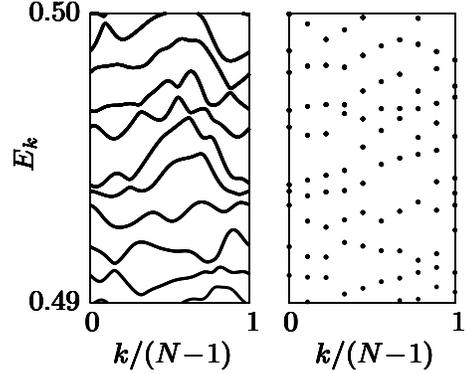}
\caption{Upper section of the $E_k$-vs.-$k$ diagram for $N =
\infty$ (continuous lines) and $N = 10$ (dots). In the limit of
small $N$ the band structure of distinct smooth lines apparently
breaks down towards a disconnected set of points. Other
parameters: $n=500$, $\delta \epsilon=0.5$, $\lambda = 0.0005$.}
\label{figurespaghetti}
\end{figure}

Standard solid state theory always predicts ballistic transport
for a translational invariant model without particle-particle
interactions. Nevertheless, in the limit of many bands (many
orbitals per site) and few sites (few $k$-values) two features may
occur: Firstly, the band structure in $k$-space becomes a
disconnected set of points rather than the usual set of distinct
smooth lines, see Fig.~\ref{figurespaghetti}. It is therefore
impossible to extract velocities by taking derivatives of
dispersion relations. And secondly, the eigenstates of the current
operator no longer coincide with the Bloch eigenstates of the
Hamiltonian such that the current becomes a non-conserved
quantity, even in the absence of impurity scattering. It is
straightforward manner to check that both features occur in the
regime where condition (\ref{condition1}) is fulfilled. This is
the regime where standard solid state theory breaks down due to
the fact that the system is too ``small''.

%
%

\subsubsection*{Validity of the TCL Approach and Failure of the NZ Technique}
Lowest order TCL suggests the emergence of ballistic transport, if
either condition (\ref{condition1}) or (\ref{condition2}) is
violated, e.~g., if the coupling strength $\lambda$ is
sufficiently weak or strong, respectively. Although the agreement
of these predictions with numerics surely is evident, it is
desirable to support them by an independent analytical
calculation, too. Such calculations are indeed possible in the
limit of $\lambda \rightarrow 0$ and $\lambda \rightarrow \infty$,
since then first order perturbation theory can be invoked:
$\lambda \, \hat{V}$ is a small perturbation to $\hat{H}_0$ in the
first case and vice versa in the second case. In both limits a
localized initial state $\psi(0) = | \mu , i \rangle$ eventually
leads to a variance $\sigma^2(t) \propto t^2$ in the limit $N
\rightarrow \infty$. The corresponding proof essentially requires
the use of Bessel functions and their properties, see Appendix
\ref{analytically} for details.

However, in a sense the agreement of the numerical simulations
with the TCL2 result is really surprising: The fact that the
correlation function $C(\tau)$ features full revivals at multiples
of $T$ points towards strong memory effects,
cf.~Fig.~\ref{figureC}. It appears to be a widespread belief that
long memory times have to be treated by means of the NZ projection
operator technique. Whereas the solutions of NZ2 and TCL2 are
almost identical for $\tau_R < T$ in the diffusive regime, for
$\tau_R \gg T$ in the deep ballistic regime the NZ2 equation
contrary predicts a purely oscillating behavior of the
corresponding density profiles $p_q(t)$. But such a behavior
obviously contradicts the observed Gaussian decay in
Fig.~\ref{figuredecay2} and consequently demonstrates the failure
of NZ2 in the description of the long-time dynamics.

\begin{figure}[htb]
\centering
\includegraphics[width=0.7\linewidth]{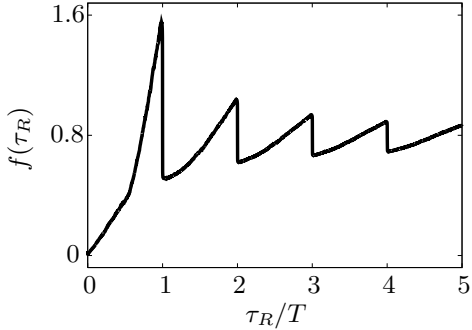}
\caption{Ratio $f(t)$ of fourth to second order contribution at $t
= \tau_R$ (``best case'' approximation). The fact that $f(\tau_R)
\sim 1$ is already reached for rather small $\tau_R \sim T$ seems
to indicate the breakdown of the TCL2 prediction for ballistic
behavior on large length scales. Parameters: $n=500$, $\delta
\epsilon=0.5$.} \label{figureratio}
\end{figure}

Unfortunately, for this model the unrestricted validity of lowest
order TCL cannot be prognosticated by the ideas of
Secs.~\ref{extension} and \ref{range}. According to
Fig.~\ref{figureratio}, at $\tau_R \sim T$ the fourth order takes
on the same order of magnitude as the second order. This finding
wrongly indicates the breakdown of the TCL2 prediction for $\tau_R
\gg T$. One is apparently concerned with a situation where higher
order contributions are not individually small but otherwise
compensate each other to approximately zero. In such a situation a
large but finite fourth order term alone is not suitable as an
``alarm'' criterion. Note that for this model the fourth order
term does not increase arbitrarily and converges to a finite
value, see Fig.~\ref{figureratio}. This will not be the case in
the following Sec.~\ref{anderson}.

%
%

\subsection{Anderson Model}
\label{anderson}

Since it had been suggested by P.~W.~Anderson, the Anderson model
served as a paradigm for transport in disordered systems
\cite{anderson1958,abouchacra1973,lee1985,kramer1993,erdos2007,schwartz2007}.
In its probably simplest form the Hamiltonian may be written as
\begin{equation}
\hat{H} = \sum_{\bf r} \epsilon^{}_{\bf r} \, \hat{a}^\dagger_{\bf
r} \, \hat{a}^{}_{\bf r} + \sum_{\text{NN}} \hat{a}^\dagger_{\bf
r} \, \hat{a}^{}_{\bf r'} \, , \label{andersonH}
\end{equation}
where the $\hat{a}_{\bf r}$, $\hat{a}^\dagger_{\bf r}$ are the
usual annihilation, respectively creation operators; $\bf r$
labels the sites of a $d$-dimensional lattice; and NN indicates a
sum over nearest neighbors. The $\epsilon_{\bf r}$ are independent
random numbers, e.~g., Gaussian distributed numbers with mean
$\langle \epsilon_{\bf r} \rangle = 0$ and variance $\langle
\epsilon_{\bf r} \, \epsilon_{\bf r'} \rangle = \delta_{{\bf r},
{\bf r}'} \, \sigma^2$. Thus, the first sum in (\ref{andersonH})
describes a random on-site potential and hence disorder.

It is well known that in the presence of disorder, $\sigma \neq
0$, the eigenstates of the Hamiltonian (\ref{andersonH}) are no
longer given by Bloch functions: The eigenstates are not
necessarily extended over the whole lattice and can become
localized in configuration space, i.~e., the envelope of a
wavefunction decays exponentially on a finite localization length.

This phenomenon and its impact on transport have intensively been
studied for the Anderson model
\cite{anderson1958,abouchacra1973,lee1985,kramer1993}. For the
lower dimensional cases, $d = 1$ and $d = 2$, it is commonly
assumed that all eigenstates feature finite localization lengths
for arbitrary (non-zero) values of $\sigma$. Consequently, in the
thermodynamic limit, i.~e., with respect to the infinite length
scale, an insulator is to be expected, see
Ref.~\cite{lee1985,kramer1993}. Of particular interest is the
$3$-dimensional case, of course. Here, the mobility edge arranges
spatially localized and extended wavefunctions into separated
regimes in energy space. When the amount of disorder is increased,
the mobility edge goes above the Fermi level and a
metal-to-insulator transition is induced at zero temperature
\cite{lee1985,kramer1993}, still on the infinite length scale.
When $\sigma$ is further increased, above some critical disorder
$\sigma_C$, all eigenstates become localized and an insulator is
to be expected for $T > 0$ also. ($\sigma_C = W_C/\sqrt{12}
\approx 6$, where $W_C$ is the critical ``width'' of the Gaussian
distribution \cite{kramer1993}.)

However, with respect to finite length scales the following
transport types are generally expected: ({\bf i}.)~ballistic on a
scale below some, say, mean free path; ({\bf ii}.)~possibly
diffusive on a scale above this mean free path but below the
localization length; and ({\bf iii}.)~insulating on a scale above
the localization length.

In the present Section, other than most of the pertinent
literature, we do not focus on the mere existence of a finite
localization length. Instead we rather concentrate on the size of
the intermediate regime and the dynamics within. We especially
demonstrate that there exists a length scale regime in which the
dynamics is indeed diffusive and characterized by an energy-independent
diffusion constant. In principle, this regime could be very large
for long localization lengths. But the results in this Section
indicate that it is not. Investigations in this direction (but not for
$d=3$) are also performed in Refs.~\cite{schwartz2007,lherbier2008}.

Our approach is still based on the form of the TCL technique which
has been established in Secs.~\ref{projection} - \ref{range}. In
this form the TCL method is restricted to the limit of infinite
temperature. This limitation implies that energy dependences are
not resolved, i.~e., our results are to be interpreted as results
for an {\it overall} behavior of all energy subspaces. Thus,
the regime in which TCL2 holds is characterized by the fact that
the dynamics within is diffusive at {\it all} energies with a
{\it single} diffusion constant.

In principle the formalism is also applicable in the case of
finite temperatures, but solely in the limit of weak coupling
strengths. In this limit the energy subspaces of the full system
are directly known from the spectra of the local subunits. The
contributions of the initial condition $p_q(0)$ to energy
subspaces can therefore be weighted with a Boltzmann factor and
may be treated separately from each other. But for strong
interactions, as it is the typical case for the Anderson model,
energy subspaces are simply not extractable from the uncoupled
system and merely the case of infinite temperature is accessible.

\begin{figure}[htb]
\centering
\includegraphics[width=0.7\linewidth]{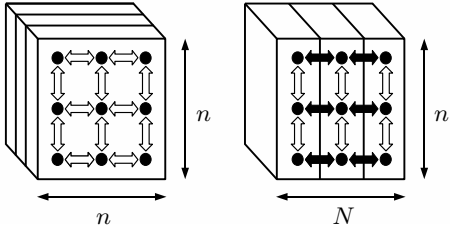}
\caption{A $3$-dimensional lattice which consists of $N$ layers
with $n \times n$ sites each. Only next-neighbor hoppings are
taken into account. Constants for intra-layer hoppings are set to
$1$ (white arrows), inter-layer hoppings are specified by another
constant $\lambda$ (black arrows). } \label{figuremodel2}
\end{figure}

As shown in Fig.~\ref{figuremodel2}, we consider a $3$-dimensional
(cubic) lattice consisting of $N$ layers with $n \times n$ sites
each. For technical reasons we use a Hamiltonian which is almost
identical to (\ref{andersonH}) with a single exception: All those
terms corresponding to hoppings between layers are multiplied by
some constant $\lambda$. However, for $\lambda=1$ the Hamiltonian
reduces to the standard Anderson Hamiltonian (\ref{andersonH}).

A ``coarse-grained'' description in terms of subunits is
established now: At first we take all those terms of the
Hamiltonian which only contain the sites of the $\mu$th layer in
order to form the local Hamiltonian $h_\mu$ of the subunit $\mu$.
Thereafter all those terms which contain the sites of neighboring
layers $\mu$ and $\mu+1$ are selected in order to form the
interaction $\lambda \, \hat{v}_{\mu,\mu+1}$ between adjacent
subunits $\mu$ and $\mu+1$. Then the total Hamiltonian may be also
written in the form of (\ref{hamiltonian}) as $\hat{H} = \hat{H}_0
+ \lambda \, \hat{V}$. Note that in this form the additional
parameter $\lambda$ allows for the independent adjustment of the
interaction strength. The eigenbasis of $\hat{H}_0$ may be found
from the diagonalization of disconnected layers.

By $\hat{p}_\mu$ we denote the particle number operator of the
$\mu$th subunit, i.~e., the sum of $\hat{a}^\dagger_{\bf r} \,
\hat{a}^{}_{\bf r}$ over all ${\bf r}$ of the $\mu$th layer. Since
the overall number of particles is conserved, $[ \, \sum_\mu
\hat{p}_\mu, \hat{H} \, ] = 0$, and no particle-particle
interactions are taken into account, we choose to restrict the
analysis to the one-particle subspace. We may therefore implement
$\hat{p}_\mu$ by (\ref{p}) with $p^i = 1/n$. The corresponding
expectation value $p_\mu(t)$ is the probability for locating the
particle somewhere within the $\mu$th subunit. The consideration
of these ``coarse-grained'' probabilities corresponds to the
investigation of transport along the direction which is
perpendicular to the layers, cf.~Fig.~\ref{figuremodel2}. Instead
of simply characterizing whether or not there is transport at all,
we analyze the full dynamics of the $p_\mu(t)$.

According to Sec.~\ref{projection}, the application of our method
requires that two conditions are fulfilled: ({\bf i}.)~the overall
conservation of probability which is naturally provided; and ({\bf
ii}.)~the ``average'' translational invariance in terms of a
correlation function
\begin{equation}
C(\tau) = \frac{1}{n^2} \, \text{Tr}\{ \, \hat{v}_{\mu, \mu+1}(t)
\, \hat{v}_{\mu, \mu+1}(t_1) \, \}
\end{equation}
which depends only negligibly on the concrete choice of the layer
number $\mu$ (during some relevant time scale). Simple numerics
indicates that this assumption is well fulfilled (for the values
of $\sigma$ which are discussed here), once the layer sizes exceed
ca.~$30 \times 30$. Exploiting this assumption immediately leads
to the TCL2 prediction (\ref{tcl2}), namely, the relaxation of
Fourier modes $p_q(t)$ with the standard rate $-W \, R(t)$. The
underlying correlation function is the above $C(\tau)$, of course.

Direct numerics shows that $C(\tau)$ again looks like a standard
correlation function. We thus retain the former notation of
$\tau_C$ as the correlation time and $R$ as the area under the
initial peak of $C(\tau)$. The numerical results also indicate
that neither $\tau_C$ nor $R$ depend substantially on $n$ (at
least for $n > 30$). Consequently, both $\tau_C$ and $R$ are
essentially functions of $\sigma$.

\begin{figure}[htb]
\centering
\includegraphics[width=0.8\linewidth]{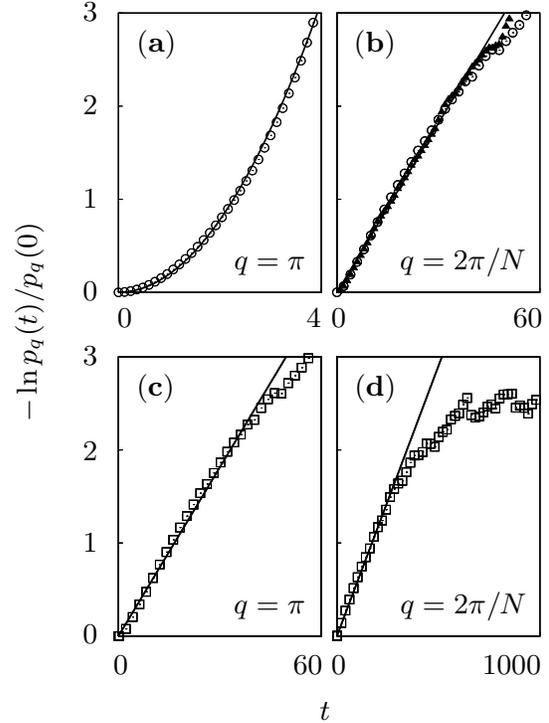}
\caption{Time evolution of modes $p_q(t)$ with $q = \pi$ (shortest
wavelength) and $q = 2 \pi / N$ (longest wavelength) for $\lambda
= 0.24$ (circles) and $\lambda = 0.08$ (squares). Remaining model
parameters: $n = 30$, $\sigma = 1$, and $N = 10$. Symbols are
obtained numerically from exact diagonalization, whereas all
curves correspond to the theoretical TCL2 predictions. Additional
data from a 4th order Suzuki-Trotter integrator is shown in ({\bf
b}) for the case of $\lambda = 1$ and $N = 42$ (triangles). Note
that the curves in ({\bf b}) and ({\bf c}) are identical, since
the factor $W$ is the same, see text for details. Further note
that the transition from ({\bf a}) towards ({\bf d}) is to be
expected for larger $N$ with a single choice of $\lambda$. For
evidence in that direction, see Fig.~\ref{figureseries}.}
\label{figure4panels}
\end{figure}

\begin{figure}[htb]
\centering
\includegraphics[width=0.5\linewidth]{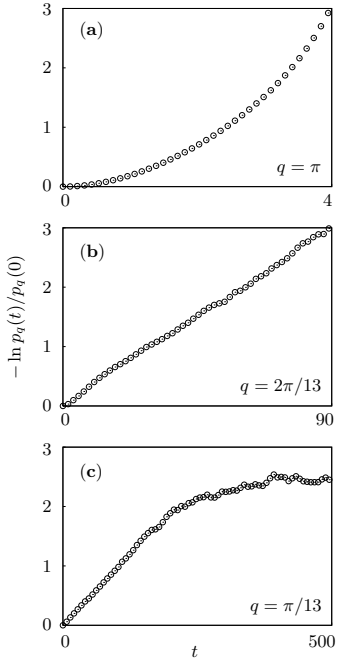}
\caption{Numerical illustration of the transition in
Fig.~\ref{figure4panels} for a fixed $\lambda$. Model parameters:
$n = 20$, $\sigma = 1$, $N =26$, and $\lambda = 0.24$. Smaller
layer sizes ($n = 20$) are chosen to allow for exact
diagonalization of a system with a larger number of layers ($N =
26$). Note that theoretical curves are not indicated, since for $n
= 20$ the ``average'' translational invariance is not well
fulfilled yet.} \label{figureseries}
\end{figure}

According to all above findings and the reasoning in
Sec.~\ref{projection}, for $t > t_C$ diffusive behavior with a
diffusion constant $D = \lambda^2 \, R$ is to be expected in TCL2,
cf.~(\ref{tcl2diffusion}). Due to the independence of $R$ from
both $n$ and $N$, the pertinent diffusion constant for arbitrarily
large systems may be quantitatively inferred from the
diagonalization of a finite, e.~g., ``$30 \times 30$ layer''.

In order to check the above theory, we exemplarily present some
results here. For $n = 30$ and, e.~g., $\sigma = 1$ we numerically
find $\tau_C \approx 10$ and $R \approx 2.9$. Thus, additionally
choosing $\lambda = 0.24$ and considering the longest wavelength
in a $N = 10$ system ($q = \pi / \, 5$), we obtain $W \, R \approx
0.064$. This corresponds to a ratio $\tau_R / \tau_C \approx 1.6$
and hence $\tau_R > \tau_C$ which justifies the replacement of
(\ref{tcl2}) by (\ref{tcl2diffusion}). And indeed, for the
dynamics of $p_q(t)$ we get an excellent agreement of the
theoretical prediction based on (\ref{tcl2diffusion}) with the
numerical solution of the full time-dependent Schr\"odinger
equation, see Fig.~\ref{figure4panels}b. Note that this solution is
obtained by the use of exact diagonalization. Naturally
interesting is the ``isotropic'' case of $\lambda = 1$. Keeping
$\sigma = 1$, one has to go to the longest wavelength in a $N =
42$ system in order to keep the $W$ of the former example
unchanged. If our theory applies, the decay curve should be the
same. This indeed turns out to hold, see
Fig.~\ref{figure4panels}b. Note that the integration in this case
already requires approximative numerical integrators like, e.~g.,
Suzuki-Trotter decompositions \cite{steinigeweg2006-1}. A
numerical integration of systems with larger $N$ rapidly becomes
unfeasible but an analysis based on (\ref{tcl2diffusion}) may
always be performed.

\begin{figure}[htb]
\centering
\includegraphics[width=0.9\linewidth]{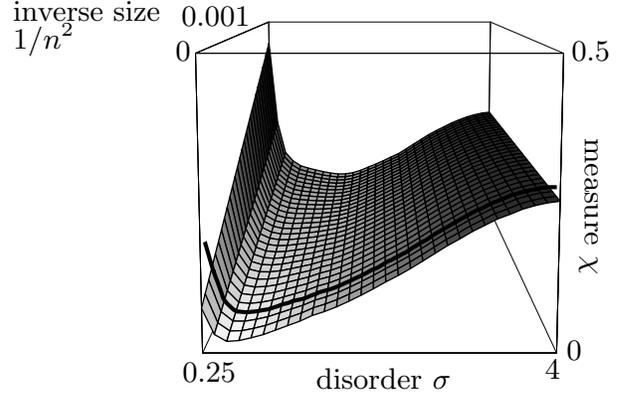}
\caption{Numerical results for the measure $\chi$ with respect to
the disorder $\sigma$ and the inverse layer size $1/n^2$ (``best
case approximation''). $\chi$ detects the corridor of length
scales where the {\it overall} behavior of all energy regimes is
diffusive. For those values of $\chi$ which are in the order of
$1/4$ the corridor does not exist. But for those values of $\chi$
which are closer to $0$, the corridor opens and diffusive behavior
is present in {\it all} energy regimes with a {\it single}
diffusion constant. The smaller $\chi$, the larger is this
corridor of diffusive length scales. An absolute minimum
$\chi_{\text{min}} \approx 0.02$ is found at $\sigma \approx 0.5$
in the limit of $n \rightarrow \infty$. Note that only $10 \%$ of
the whole area is extrapolated (the area in front of the thick
line).} \label{measure}
\end{figure}

So far, we have characterized the dynamics of the diffusive
regime. We now turn towards an investigation of its size. To this
end we consider the measure $\chi$ which has been introduced in
Secs.~\ref{extension} and \ref{range}. Recall that this measure
yields $l_\text{min} / l_\text{max} \approx 2 \, \sqrt{\chi}$,
cf.~(\ref{rangel}).

$l_\text{min}$ is the length scale where diffusive dynamics breaks
down due to the fact that the corresponding $p_q(t)$ decays on a
comparatively short time scale $\tau_R \approx \tau_C$. In
complete analogy to the former model, this transition is still
correctly described by TCL2: $R(t)$ is not approximately constant
but linearly increases during the relaxation period. This strongly
indicates a transition towards ballistic behavior, see
Fig.~\ref{figure4panels}a.

Contrary, $l_\text{max}$ is the length scale where diffusive
dynamics breaks down, because the corresponding $p_q(t)$ decays on
a long time scale $\tau_R$ at which the additional contribution
$S(t)$ becomes non-negligible, cf.~(\ref{S}). Since the
interaction perfectly fulfills the van Hove structure, $S(t)$
essentially reflects fourth order effects which account for,
e.~g., insulating behavior, cf.~Fig.~\ref{figure4panels}d. [Note
that all our data available from exact diagonalization is in
accord with a description based on (\ref{S}). Note further that
$S(t)$, other than $R(t)$, scales significantly with $n$. This
eventually gives rise to the $n$-dependence in
Fig.~\ref{measure}.]

As shown in Fig.~\ref{measure}, for each layer size $n$ there is
some disorder which minimizes $\chi$ and hence ``optimizes'' the
diffusive regime. However, for $n = 30$ (back of
Fig.~\ref{measure}) we find $\sqrt{\chi_\text{min}} \approx 1/3$
at this optimum disorder, indicating about one diffusive
wavelength. Exactly those respective wavelengths have been chosen
for the examples in Figs.~\ref{figure4panels}b,c but not in
Fig.~\ref{figure4panels}d. For all $\sigma$ and up to $n = 100$
(which is about the limit for our numerics) $\chi$ clearly appears
to be of the form $\chi(\sigma, n) = A(\sigma) / n^2 + B(\sigma)$.
Extrapolating this $1/n^2$-behavior yields a suggestion for the
infinite model $n = \infty$ (front of Fig.~\ref{measure}).
According to this suggestion, we find $\sqrt{\chi_\text{min}}
\approx 1/7$, again at optimum disorder. This indicates a rather
small regime of diffusive wavelengths, even for the infinite
system.

We finally recall that these findings apply at infinite
temperature, i.~e., the small diffusive regime is characterized by
the fact that the dynamics within is diffusive at {\it all}
energies with a {\it single} diffusion constant. The absence of
diffusive dynamics in the limit of strong disorder apparently
agrees with the common expectation that diffusion at each energy
stops completely, once $\sigma$ reaches a value larger than
$\sigma_C \approx 6$ and all eigenstates become localized
\cite{lee1985,kramer1993}. Down to intermediate amounts of
disorder the localization phenomenon certainly plays a crucial
role for the smallness of the diffusive regime. In the limit of
weak disorder, however, localization appears to be less important,
especially since only few eigenstates in the outer tails of the
density of states are localized, while the overwhelming majority
is extended. In that limit the smallness of the regime in which TCL2
holds may result from the fact that diffusive dynamics is present at
each energy but with a energy-dependent diffusion constant,
cf.~\cite{kramer1993}. Then the overall behavior of all energy
subspaces is a multi-exponential decay and not diffusive,
of course. On that account it appears plausible that the diffusive
regime becomes much larger, if only a certain subspace of energy
is considered.

%
%

\section{Summary and Conclusion}
\label{summary} In this work the TCL projection operator technique
has been applied to quantum transport in modular systems. In
particular, the projection onto local densities and the lowest
order contribution of the TCL expansion have been used as a
strategy for the analysis of transport and its length scale
dependence. Furthermore, an estimation for the range of validity
of the lowest order prediction has been derived by means of an
extension of the standard projection that includes additional
relevant degrees of freedom. This estimation is especially suited
for interaction types with van Hove structure and also provides an
interpretation in the context of length scales.

As a first application a single-particle model has been
investigated which has randomly structured interactions but
nevertheless is translational invariant. For this model a full
characterization of all types of transport behavior has been
obtained. Remarkably, purely diffusive behavior has been
demonstrated to occur on an intermediate length scale which is
bounded by completely ballistic regimes in the limit of short and
long length scales.

The next step in the context of single-particle models has been
performed by the introduction of disorder. The 3D Anderson model
has been shown to exhibit fully diffusive dynamics on an
intermediate length scale between mean free path and localization
length. But it has also been demonstrated that the diffusive
regime is extremely small, at least in the considered limit of
high temperatures.

Naturally, a further important step is given by the application of
the theory to interacting many-particle systems with or without
disorder. In fact, there are already preliminary results for a
``many-particle'' model which is translational invariant, has
randomly structured interactions but does not allow for any
single-particle restriction. This model has been discussed only
briefly \cite{steinigeweg2006-2,gemmer2006} and basically seems to
show the same dynamical behavior as its single-particle analog,
that is, diffusive dynamics on intermediate length scales breaks
down towards ballistic transport in the limit of large length
scales. The main problem for this system arises from the fact that
a direct comparison of the second order prediction with the
numerical solution of the time-dependent Schr\"odinger equation is
not feasible for sufficiently large systems. Consequently, an
appropriate estimate for the validity range of second-order TCL of
the type derived in this paper is indispensable.

%
%

\begin{appendix}

%
%

\section{Analytically Solvable Limit Cases}
\label{analytically}

The results of Sec.~\ref{random} strongly suggest the occurrence
of ballistic transport in the two limit cases (\ref{condition1})
and (\ref{condition2}) of weak and strong coupling strength
$\lambda$. Since these results depend on the applicability of the
TCL2 method, it would be desirable to support them by independent
analytical calculations. Such calculations are possible in the two
limit cases $\lambda \rightarrow 0$ and $\lambda \rightarrow
\infty$, since then first order perturbation theory can be
invoked. Moreover, we will restrict ourselves to the case where
the Hamiltonian given by Eqs.~(\ref{hamiltonian})-(\ref{v}) is
translational invariant and the next-neighbor interaction matrix
(\ref{v}) is symmetric, i.~e., we will assume
\begin{eqnarray}
&& h_\mu^i = h_0^i \equiv h^i \; , \label{ass_h} \\
&& v_{\mu,\mu+1}^{i,j} = v_{0,1}^{i,j}= v_{0,1}^{j,i} \equiv
v^{i,j} \; . \label{ass_v}
\end{eqnarray}
In these calculations various sums over $\mu = 0, 1, \ldots, N-1$
occur, which will be most conveniently approximated by integrals.
This approximation is exact in the limit $N \rightarrow \infty$.
The variance (\ref{variance}) will be considered with local
``excitation densities" (\ref{p}) using $p_\mu^i = 1$.
\\
Let us assume that the symmetric next-neighbor interaction matrix
$v$ has been diagonalized:
\begin{equation}
\sum_{j=0}^{n-1} v^{i,j} \, S_k^j = s_k \, S_k^i \label{diag_v}
\end{equation}
We will consider the time evolution of an initially localized
excitation, i.~e., a solution
\begin{equation}
\varphi(t) = \sum_{\mu, i} \varphi_{\mu}^i(t) \, | \mu, i \rangle
\label{sol_phi}
\end{equation}
of the Schr\"odinger equation with the initial value
$\varphi_\mu^i(0) = \delta_{\mu,0} \, \delta_{i,0}$.

\subsubsection{The case $\lambda\rightarrow 0$}

In the limit $\lambda \rightarrow 0$ and $N \rightarrow \infty$ we
obtain
\begin{equation}
\left | \varphi_\mu^i(t) \right |^2 \sim \delta_{i,0} \left |
J_{|\mu|} \! \left ( 2 \, \lambda \, v^{0,0} \, t \right ) \right
|^2 \; , \label{lambda_0_1}
\end{equation}
where $J_{|\mu|}$ denotes the $|\mu|$-th Bessel function, and as a
consequence
\begin{equation}
\sigma^2(t) = \sum_{\mu, i} \mu^2 \, \left | \varphi_\mu^i(t)
\right |^2 \sim 2 \, \lambda^2 \, v^{0,0} \, t^2 \, ,
\label{lambda_0_2}
\end{equation}
indicating ballistic transport.

\subsubsection{The case $\lambda \rightarrow \infty$}

Upon rescaling the Hamiltonian (\ref{hamiltonian}) in the form
\begin{equation}
\hat{H} = \frac{1}{\lambda} \,\hat{H_0} + \hat{V}
\label{lambda_infty_rescale}
\end{equation}
we can again apply first order perturbation theory. This yields
eigenvalues of the form
\begin{equation}
E_{k, \mu} = \frac{1}{\lambda} \, E_k + 2 \, s_k \cos \frac{2 \pi
\, \mu}{N} \, , \label{lambda_infty_eigenvalues1}
\end{equation}
where
\begin{equation}
E_k = \sum_i \left| S_k^i \right|^2 h^i \, .
\label{lambda_infty_eigenvalues2}
\end{equation}
Consequently,
\begin{eqnarray}
&& \!\!\! \left | \varphi_\mu^i(t) \right|^2 \nonumber \\
\sim && \!\!\! \left | \sum_k \overline{S_k^0} \, S_k^i \, \exp
\left ( \frac{-\imath \, E_k \, t}{\lambda} \right ) J_{|\mu|} \!
\left( 2 \, s_k \, t \right ) \right |^2 \! .
\label{lambda_infty_phi_1}
\end{eqnarray}
In order to evaluate the variance (\ref{variance}) we will utilize
the following formula, stated without proof,
\begin{eqnarray}
&& \!\!\! \sum_{m = -\infty}^\infty \, m^2 \, J_{n+m}(s) \, J_{m}(t) \nonumber \\
= && \!\!\! \frac{t^2}{4} \left [ J_{n+2}(s-t) + J_{n-2}(s-t) + 2
\, J_n(s-t)
\right ] \nonumber \\
+ && \!\!\! \frac{t}{2} \left [ J_{n+1}(s-t) - J_{n-1}(s-t) \right
] \label{lambda_infty_J}
\end{eqnarray}
in the special case of $n=0$. We then obtain for the leading term
\begin{equation}
\sigma^2(t) \sim 2 \, t^2 \, \sum_k \left | S_k^0 \right |^2 s_k^2
+ {\cal O}(t) \; , \label{lambda_infty_var}
\end{equation}
again indicating ballistic transport. Numerical tests show that
the parabolic approximation (\ref{lambda_infty_var}) is very close
for all times.

\end{appendix}

%
%

\begin{acknowledgement}

We sincerely thank C.~Bartsch and M.~Kadiro\={g}lu for fruitful
discussions. Financial support by the Deutsche
Forschungsgemeinschaft is gratefully acknowledged. One of us
(HPB) gratefully acknowledges a Fellowship of the
Hanse-Wissenschaftskolleg, Delmenhorst.

\end{acknowledgement}

%
%

\end{document}